# Delusion and Deception in Large Infrastructure Projects

## Two Models for Explaining and Preventing Executive Disaster*

**Bent Flyvbjerg**

Corresponding author

Aalborg University

Dept. of Development and Planning

flyvbjerg@plan.aau.dk

and

Delft University of Technology

Faculty of Technology, Policy, and Management

**Massimo Garbuio**

University of Western Australia

UWA Business School

massimo@biz.uwa.edu.au

**Dan Lovallo**

University of Sydney

D.Lovallo@econ.usyd.edu.au

Bent Flyvbjerg, Massimo Garbuio, and Dan Lovallo, "Delusion and Deception in Large Infrastructure Projects: Two Models for Explaining and Preventing Executive Disaster," *California Management Review*, vol. 51, no. 2, Winter 2009, pp. 170-193.

DOI: 10.1225/CMR423

Link to published article:

www.emeraldinsight.com/bibliographic_databases.htm?id=1791300&show=abstract

*The authors would like to thank Sara Beckman and Nuno Gil as well as three anonymous reviewers for their helpful comments.

*The Economist* recently reported that infrastructure spending is the largest it is ever been as a share of world GDP. With $22 trillion in projected investments over the next ten years in emerging economies alone, the magazine calls it the "biggest investment boom in history."[1] The efficiency of infrastructure planning and execution is therefore particularly important at present. Unfortunately, the private sector, the public sector and private/public sector partnerships have a dismal record of delivering on large infrastructure cost and performance promises. Consider the following typical examples.

*In January 2003, Toll Collect – a consortium of DaimlerChrysler, Deutsche Telekom, and Cofiroute of France – was scheduled to start tolling heavy trucks on German motorways for the Federal government. The new tolling system was designed to be a showcase for public-private partnership in infrastructure management. A year later the project was falling apart. The developers had been too optimistic about the software that would run the system. The government was losing toll revenues of €156 million ($244m) a month, caused by delays, and estimated to total €6.5 billion before problems could be fixed. For lack of funds, all new road projects in Germany and related public works were put on hold, threatening 70,000 construction jobs. Politicians and members of the media were calling for prosecution of Toll Collect for deceiving the government. Finally, the German transport minister cancelled the contract with Toll Collect and gave the company two months to come up with a better plan, including how to fill the revenue shortfall. By the time tolling at last started, after further delays in 2005, "Toll Collect" had become a popular byword among Germans used to describe everything wrong with the national economy.*

*In 1987, Eurotunnel, the private company that operates the tunnel under the English Channel, went public to raise funds for the project. The Channel tunnel was*

*prime minister Margaret Thatcher's flagship project to show the world how private business could effectively provide public infrastructure. Eurotunnel told investors that building the tunnel--the longest of its kind in Europe--would be relatively straightforward and that 10% "would be a reasonable allowance for the possible impact of unforeseen circumstances on construction costs."[2] Once built, the real cost of the project was double the forecasted costs in constant dollars. Initially, the misinformation about costs and risks served the purpose of getting the project started. From the 1987 IPO until cost overruns hit the project one and a half years later, share prices more than tripled. Then they fell by two thirds and, when it became clear that revenue projections were as biased as cost forecasts, by another two-third. In 1995, Eurotunnel stopped interest payments on its loans and began a decade-long, tumultuous process of financial restructuring from which it did not recover until 2007. The intended flagship of privatization became a scare story for business and set back the process of infrastructure privatization by at least a decade.*

*In 1959, the construction of Sydney Opera House started before either drawings or funds were fully available. The initial budget of seven million Australian dollars was a political, low-balled budget designed for project approval before the coming elections.[3] Eventually the Opera House was opened in 1973, 10 years later than the original planned completion date, at a cost of 102 million Australian dollars. It holds the world record for cost overrun at 1,400 percent and this was for a scaled-down version of the original design. This figure does not include 45 million dollars allocated in 2002 in part to bring the building more in agreement with the architect, Jørn Utzon's original plans.*

## Over Budget, Over Time, Over and Over Again

There are some phenomena that have no cultural bounds such as maternal love and a healthy fear of large predators. We can add to this list the fact that, across the globe, large infrastructure projects almost invariably arrive late, over-budget and fail to perform up to expectations. Cost overruns and benefit shortfalls of 50 percent are common; cost overruns above 100 percent are not uncommon. For example, in one study of major projects in 20 countries, nine out of ten projects had cost overruns.[4] Similarly, a study of 44 urban rail projects in North America, Europe, and developing nations, including London's Tube and the metros in Washington DC and Mexico City, found that the average construction cost overrun in constant prices was 45 percent; for a quarter of the projects cost overruns were at least 60 percent. In addition, passenger ridership was, on average, 50 percent lower than forecast. Furthermore, for a quarter of the projects, ridership was at least 70 percent lower than estimated.[5] An appropriate slogan seems to be "over budget, over time, over and over again". As comforting as it is to know that we are not alone in our folly, it would be even better to minimize the gap between expectations and performance for projects that consume such a large share of the private and, especially, public purse.

Executives typically attribute project underperformance to numerous uncertainties such as project complexity, technological uncertainty, demand uncertainty, scope clarity, unexpected geological features and negative plurality (i.e., opposing stakeholder voices).[6] No doubt, all of these factors at one time or another contribute to cost overruns, benefit shortfalls, and time delays. The goal of this article, however, is not to explain, for example, how to implement complex projects more efficiently by over-coming these uncertainties. Rather, we explain why cost, benefits and time forecasts for more complex projects are systematically over-optimistic in the planning phase in comparison to less complex projects. In other words, "why do project planners, on average, fail to anticipate the greater costs of complex projects or those based on new technologies, etc.?"

The underlying reasons for all forecasting errors can usefully be grouped into three categories: 1) delusions or honest mistakes; 2) deceptions or strategic manipulation of information or processes or 3) bad luck.[7] Bad luck or the unfortunate resolution of one of the major project uncertainties is the attribution typically given by management for a poor outcome.[8] While not denying such a salient explanation, this article explores the underlying psychological and governance reasons for mis-estimation rather than proximate engineering causes.

Deliberately or not, risks of scope changes, high complexity, and unexpected geological features are systematically underestimated during project preparation. Both delusion and deception see the high failure rates for ventures as a consequence of flawed decision making. According to the first explanation - delusion - the flaw consists in executives falling victim to what psychologists call the planning fallacy.[9] In its grip, managers make decisions based on delusional optimism rather than on a rational weighting of gains, losses, and probabilities. They overestimate benefits and underestimate costs and time. They involuntarily spin scenarios of success and overlook the potential for mistakes and miscalculations. As a result, managers pursue initiatives that are unlikely to come in on budget or on time, or to ever deliver the expected returns. These biases are often the result of the *inside view* in forecasting: decision makers have a strong tendency to consider problems as unique and thus focus on the particulars of the case at hand when generating solutions.[10] Adopting an *outside view* of the problem has been shown to mitigate delusion. It is applied by ignoring the specific details of the project at hand and uses a broad reference class of similar projects to forecast outcomes for the current project.

According to the second explanation - deception - decision-making is flawed by strategic misrepresentation or the presence of what economists refer to as principal-agent problems. Whereas the first explanation is psychological, the second is due to the different preferences and incentives of the actors in the system.[11] In this situation, politicians, planners or project champions deliberately and strategically overestimate benefits and underestimate costs in order to increase the likelihood that their projects, and not their competition's, gain approval and funding. These actors purposely spin scenarios of success and gloss over the potential for failure. This results in managers promoting ventures that are unlikely to come in on budget or on time, or to deliver the promised benefits. However, this misrepresentation and failure can be moderated by measures that enhance transparency, provide accountability, and align incentives.

While delusion and deception have each been addressed in the management literature before, in what follows they are jointly considered for the first time. In addition, they are specifically applied to infrastructure problems in such a way that both academics from diverse fields and, more importantly, practitioners can understand and implement the suggested corrective procedures. Among its contributions, the article provides a framework for analyzing the relative explanatory power of delusion and deception in such a way that it is possible to disentangle whether non-accurate forecast are more likely to be due to one or the other explanation, or both. Moreover, it suggests a simplified framework for analyzing the complex principal-agent relationships that are involved in the approval and construction of large infrastructure projects. This will facilitate the design of incentive systems and corrective procedures for improving forecasts, which is a major goal of the article.

The article is organized in two main parts. In the first part, we initially discuss the cognitive factors that produce delusion and then political and economic explanations of deception. An analysis of the complementary nature of explanations based on delusion and deception follows. In the second part, we build on these models and show how appropriate financial and non-financial incentives can mitigate deception. To conclude, we illustrate reference class forecasting, an outside view debiasing technique that has proven successful in overcoming both delusion and deception in private and public investment decisions.

## Delusion and Deception in Large Capital Projects

### Delusion

Our first explanation – delusion – accounts for the cost underestimation and benefit overestimation that occurs when people generate predictions using the inside view. Executives adopt an *inside view* of the problem by focusing tightly on the case at hand, by considering the plan and the obstacles to its completion, by constructing scenarios of future progress, and by extrapolating current trends.[12] In other words, by using typical bottom-up decision making techniques, they think about a problem by bringing to bear all they know about it, with special attention to its unique details. Below we illustrate two cognitive delusions the inside view facilitates: the planning fallacy as well as a heuristic - rule of thumb - called anchoring and adjustment.

When forecasting the outcomes of risky projects, executives often fall victim to the *planning fallacy*. Psychologists have defined it as the tendency to underestimate task-completion times and costs, even knowing that the vast majority of similar tasks have run late or gone over budget.[13] It is a well-established bias in the experimental literature. In one set of experiments, Buehler, Griffin and Ross assessed the accuracy of psychology students' estimates of completion times for their year-long honors thesis project.[14] In the experiments, the students' "realistic" predictions were overly optimistic: 70% took longer than the predicted time, even though the question was asked toward the end of the year.

On average, students took 55 days to complete their thesis, which was 22 days longer than predicted. Similar results have been found with various types of subjects and for a wide variety of tasks such as holiday shopping, filing taxes, and other routine chores.[15]

These findings are not limited to the laboratory. Cost and time overruns in large infrastructure projects have been studied by considering numerous contractual arrangements. In the case of conventional procurement, in which the public entity separately engages with several private companies, each of them providing a specific part of the service, costs and times overruns have been systematically observed in a wide range of projects.[16] In business, executives and entrepreneurs seem to be highly susceptible to this bias. Studies that compared the actual outcomes of capital investment projects, mergers and acquisitions, and market entries with managers' original expectations for those ventures show a strong tendency towards overoptimism.[17] An analysis of start-up ventures in a wide range of industries found that more than 80% failed to achieve their market-share target.[18]

*Anchoring and adjustment* is another consequence of the inside view thinking that leads to optimistic forecasts.[19] Anchoring on plans is one of the most robust biases of judgment. The first number that is considered as a possible answer to a question serves as an "anchor". Even when people know that the anchor is too high or too low, their adjustments away from it are almost always insufficient. A classic experiment revealed the power of anchoring and insufficient adjustment. People were asked to estimate various percentages, such as the percentage of African countries in the United Nations.[20] For each quantity, a number was determined by spinning a wheel of fortune in the presence of the subject. The subjects were first asked to indicate whether the number was higher or lower than the percentage of African countries and then to estimate the percentage by moving upward or downward from the arbitrary number. The arbitrary number had a substantial effect on the estimates. For instance, the median estimate of the percentage of African countries in the United Nation was strongly related to the starting points: individuals who received 10, estimated 25% whereas those that received 65, estimated 45%. These subjects started from a random anchor and then insufficiently adjusted away from it.

Similar results have been found with experienced real estate brokers who were asked to assess the value of a property.[21] These agents unanimously agreed that they did not factor a house's listing price into their evaluation of its "true" value. Each of the agents was given a 10 page booklet on the house that was being sold, which included information specific to the house as well as information about the prices and characteristics of other houses in the area that had recently been sold. The only difference in the information that the various brokers received was the listing price of the house, which was randomly manipulated within a range of plus or minus 11% of the actual listing price. The agents then went out and visited the house that was being sold, as well as several other houses in the neighborhood. The listing price significantly affected these experienced agents' evaluations. Furthermore, when told about the results, the agents maintained that the listing price anchor had no effect!

In the context of planning for a large infrastructure project there is always a plan, which is very likely to serve as an anchor. Furthermore, the plan that is developed is almost always seen as a "realistic" best or most likely case. Executives know that events may develop beyond the best or most likely case so they generally attempt to capture unforeseen costs by building in a contingency fund that is proportional to the size of the project (e.g. for cost overruns in capital investment projects). However, when compared with actual cost overruns, such adjustments are clearly and significantly inadequate.[22] Furthermore, the initial estimate serves as an anchor for later stage estimates, which never sufficiently adjust to the reality of the project's performance.

The power of these heuristics and biases is well illustrated in a field study where the Rand Corporation examined 44 chemical process plants (Pioneer Process Plants), owned by firms such as 3M, du Pont and Texaco among others. Actual construction costs were over twice as large as the initial estimates.[23] Furthermore, even a year after start-up about half of the plants (21) produced at less than 75% of their design capacity, with a quarter of the plants producing at less than 50% of their design capacity. Many of the plants in this latter category had their performance expectations permanently lowered. As illustrated in Figure 1, the typical initial estimate is less than half the final cost. Furthermore, at every subsequent stage of the process, managers underestimate the cost of completing the construction of Pioneer Process Plants.

----------------------------------

Add Figure 1 here

----------------------------------

## Deception

Our second explanatory model – deception – accounts for flawed planning in decision making in terms of politics and agency issues. With this model we introduce political and organizational pressures in executive decision-making. We describe the principal-agent problem first and the sources of strategic deception second.

### The Principal-Agent Problem in Large Capital Investments

In this section we talk about principal-agent (P-A) problems, which have mainly been examined in the context of private firms but which can be even more pernicious in public situations.[24] These are defined by relationships where a principal engages an agent to act on his or her behalf. Typical examples include a Board hiring a CEO to manage the company on behalf of the shareholders or a manager hiring an employee to carry out tasks. In fact, there is a P-A relationship for every two levels in an organization.

Large capital investment projects are situations where a multi-tier principal-agent problem exists.[25] An example will help illustrate the framework. Consider a local government that intends to build a new tunnel across a large capital city, for the benefits of the local residents and, more broadly, of the state population. The focal project will compete with other projects for funds from the state government. Once the approval is obtained, the local government puts construction out for tender. The winning bidder will carry out the construction of the infrastructure.

Figure 2 graphically represents the complexity of the P-A relationships in the case of a large capital investment proposed by a local government to the state government. In this specific example, there are three tiers of P-A relationships.

---------------------------------

Add Figure 2 here

---------------------------------

The first tier encompasses the relationship between taxpayers and the state government. Taxpayers are the principal, whereas the state government is the agent of the taxpayers that is supposed to act in their interest. As the final beneficiaries of the infrastructure, taxpayers expect projects to deliver the largest possible benefits to the community, by incurring minimal costs, attenuating risks and reaching completion within an agreed timeline. Individuals in the state government, who are elected by taxpayers, typically have their own interests, for example being re-elected and/or being remembered for the building of monumental infrastructures.

The second tier of P-A relationships has the local government acting as the agent of both taxpayers and state government. With respect to the taxpayers, the local government has the duty to propose infrastructures that provide the largest benefits to the community, and that are delivered on budget and on time. With respect to the state government, it has a duty of suggesting the best allocation of the taxpayer funds. Moreover, holding the most complete data about costs and benefits of the infrastructure that it proposes, it has the duty of providing the state government with the most accurate forecasts needed to make an informed decision. However, given the competition for scarce resources, the local government has an interest in understating its risks and costs, while overstating their benefits.

The mechanism of benefit overestimation is very simple as explained by an interviewee in research done by one of us (Flyvbjerg) and the Danish consultancy firm Cowi: "The system encourages people to focus on the benefits--because until now there has not been much focus on the quality of risk analysis and the robustness [of projects]. It is therefore important for project promoters to demonstrate all the benefits [...]."[26] In addition, knowing that the next election usually happens before the time that the proposed project is built and sometimes even approved, the local government has little interest in providing accurate forecasts. As shown by Flyvbjerg in his research on the Sydney Opera House, Joe Cahill, the Labor-premier of New South Wales, publicized a political budget for approval and fast-tracked construction to start before the elections, in case the Labor party lost the elections and attempts would be made to stop the project. According to Bob Carr, who later followed Cahill as premier of New South Wales, Cahill instructed his people "to go down to Bennelong Point [the site of the Sydney Opera House] and make such progress that no-one who succeeds me can stop this [the Opera House] going through to completion."[27] When the premier's lowballing and fast-tracking of the Opera House inevitably led to cost overruns, the architect, Jørn Utzon, was blamed. Utzon prefers to remain out of the public eye, but his son, architect Kim Utzon, explains in lieu of his father: "It was a political decision to publicize a low budget for the building, which was expected to gain approval in the political system, but which was very quickly exceeded. So even if the cost overrun turned out to be 1,400 percent in relation to the publicized budget, this budget was an eighth of the real budget for the building. So the real cost overrun is only 100 percent. The rest was politics."[28]

The third tier of P-A relationships involves the local government as the principal of agents hired to provide specific services, such as analysts and planners as well as contractors. Analysts and planners are engaged to gather the information necessary for making the final go/no-go decision. They have an incentive to provide information that is compatible with pleasing the local government, having the project approved, and being re-engaged on the next project. A manager on a large infrastructure project explained to Flyvbjerg and Cowi in their research on transport infrastructure management in the UK: "Most decent consultants will write off obviously bad projects but there is a grey zone and I think many consultants in reality have an incentive to try to prolong the life of projects, which means to get them through the business case. It is in line with their need to make a profit."[29] Another interviewee in the same study recognized that planners have better information than politicians but have no incentive to reveal such information, but rather the opposite: "You will often as a planner know the real costs. You know that the budget is too low but it is difficult to pass such a message to the counsellors [politicians] and the private actors. They know that high costs reduce the chances of national funding." Similarly, builders have the primary interests to win the tender, by offering the lowest possible price, since they know that re-contracting is often possible and, unless the contract is a fixed price and lump sum contract, delays will be tolerated. Even if interests are divergent in this case, cost-overruns and delays are tolerated unless the local government is held responsible. Clearly, the multi-tier relationship described in this specific example of the construction of a cross-city tunnel can be easily extended to the approval of any kind of large public infrastructure project.

**Sources of strategic deception**

There are certain conditions, however, that make strategic deception more likely within *each* P-A relationship. Self-interest, asymmetric information, differences in risk preferences and time horizons as well as the clarity of accountability are among the most cited causes.

A necessary condition for P-A conflicts is a difference in the actors' *self-interest*. Executive ventures, public and private, are often multimillion- and sometimes even multibillion-dollar projects. When they go forward, many stakeholders (e.g. contractors, engineers, architects, bankers, landowners, construction workers, lawyers, accountants and developers) have widely divergent incentives. In addition, politicians and executives may use ventures to jockey for position and to build monuments, which allows administrators to get larger budgets, and cities to acquire investments in infrastructure that would otherwise go elsewhere.[30] If these stakeholders are involved in, or indirectly influence, the forecasting of costs and benefits at the approval stage (the business case), this is liable to bias the entire subsequent process.

Political and economic self-interest also exists at the level of cities and states. Pickrell pointed out that transit capital investment projects in the US compete for discretionary grants from a limited federal budget each year, and that this creates an incentive for cities to make their projects look better, or else some other city may get the money.[31] Flyvbjerg and Cowi found similar results for the UK.[32] Altshuler and Luberoff, Delaney and Eckstein, and Morris and Hough found corresponding results for other project types, including major roads, tunnels and bridges, airports, stadiums, power stations, oil and gas extraction, and IT systems.[33]

A second source of strategic deception is the presence of *asymmetric information*, which means the agent who champions a project (e.g. the local government in the example above) has information that the principal does not (e.g. the state government). Being unaware of all the relevant information, the principal and ultimate decision maker may be easy to deceive. In a recent study, Flyvbjerg and Cowi interviewed public officials, planners, and consultants who had been involved in the development of large UK transportation infrastructure projects.[34] This study shows that strong interests and strong incentives exist at the project approval stage to present projects' costs and benefits as favorably as possible. Local authorities, local politicians, local officials, and some consultants (as agent's) all stand to benefit from a project that looks favorable on paper and have little incentive to actively avoid distorted estimates of benefits, costs, and risks. National bodies, like the Department for Transport and the Treasury, act as a principal and usually fund and oversee projects. They usually have an interest in more balanced appraisals, but so far they have had little success in achieving such balance. This situation may be changing with recent schemes to curb the optimism bias, which were initiated by HM Treasury in order to gain better predictability and control in public budgeting.[35]

A third relevant source of P-A issues is the presence of *different risk preferences*. For instance, if the principal is risk averse, the agent who submits a proposal for approval will have to downplay the possible risks of the venture in order to convince the principal. Because of this, managers lower down in the organization may have to misrepresent, hide and manipulate information in order to get requests for funds approved.

In addition to any inherent difference in risk preferences between decision makers and organizations or the public, another factor that comes into play is the vastly *different time horizons* the actors use to evaluate the decision. Typically, this is very long for taxpayers, but less than a decade for the individuals that are acting on their behalf. These agents may also be concerned with being remembered for initiating monumental infrastructure or, more prosaically, being re-elected. In fact, whereas a standard election term is 4 years, the average length of the time from the start of planning to start of operations for a large infrastructure project is commonly 10-15 years.[36]

Finally, another condition that leads to strategic deception is *diffuse or asymmetric accountability*. When multiple people are responsible for the ultimate success or failure of a project, it can be difficult for any one agent to be held accountable for a bad outcome. If a new initiative fails, it is often hard to place the blame squarely on one or a few actors. This lack of accountability *ex-post* can cause the agent to promote *ex-ante* projects that protect them from being held accountable if it fails, which may not be the project that maximizes the principal's total payoff.[37] In addition, the lack of clear accountability can exacerbate the problem of asymmetric information and differences in risk preferences, causing the agent to take more risk than the principal would like.

## Diagnosing the Relative Impact of Delusion and Deception

Delusion and deception are two complementary rather than alternative explanations of failure of large infrastructure projects due to cost underestimation and benefit overestimation. Although, in practice, it is often difficult clearly disentangle the two explanations. There are situations, however, where the explanatory power of one of the two models is relatively higher.

The relative strength of each explanation depends on different factors. The key to minimizing delusion is to have a good learning environment. Learning occurs "when closely similar problems are frequently encountered, especially if the outcomes of decisions are quickly known and provide unequivocal feedback".[38] Whereas, the problem of strategic deception occurs when incentives are mis-aligned. The underlying causes of these mis-alignments are differences in preferences, time horizons, incentives, and information between principals and agents.

Figure 3 describes situations where we can expect explanations due to delusion, deception, both, or neither to operate. The figure is divided in four regions. When the learning environment is good and incentives are well-aligned, there is minimal scope for delusion or deception and forecasts tend to be unbiased. Weather forecasts are an example. Meteorologists have no reason to lie and the feedback they receive is so frequent and unambiguous detailed computer models guide their predictions.

Small entrepreneurs' who own the vast majority of their companies have incentives that are well aligned so that we expect most of their errors to be due to delusions. These forecasting errors tend to be quite large. For example, 33% of entrepreneurs perceive their chances of success to be certain, which is obviously is deluded given that over 80% of such ventures fail.[39]

Many computer gaming companies release numerous titles frequently, so their learning environments are good yet they continuously state release dates they do not stick to. This type of deception has been labeled "Cheap Talk" and is designed to pre-empt sales of competitors' products.[40]

The largest errors occur when delusions and deception operate simultaneously. Even within this section useful distinctions can be made. For example, in the private sphere, we consider the construction of process plants. These need to be distinguished between normal process plants and pioneer process plants. They both have similar incentives since they are owned by the same companies, but the former is relatively more frequent and therefore learning is improved and forecasts are more accurate.

In the public sphere, we consider the examples of rails and roads. Cost underestimation and overrun for rail are on average approximately twice that for roads.[41] Similarly, average ridership overestimation for planned rail projects is around 100%, whereas such bias is not found for road projects.[42] The differences between rail and road are statistically highly significant and may largely be explained in terms of differences in incentive structures and the possibility to learn from previous and similar projects. In fact, rail projects typically compete for discretionary grants from a limited national or federal budget. This creates an incentive for promoters (agents) to make their projects look better on paper with artificially high benefit/cost ratios, or else the central government (the principal) may decide to fund some other project.[43] That is, there are incentives to provide biased estimates.

In the case of roads, funds are typically allocated as block grants with a certain amount of dollars allocated for road building as where individual projects do not compete for funds directly against each other or against other types of projects outside the highway agency. As a consequence, the misalignment of incentives between promoters and approvers are higher for rail than for roads. This conclusion is supported by a study of stated causes for inaccuracies in traffic forecasts for 234 rail and road projects. For rail projects, deception in terms of "deliberately slanted forecasts" was explicitly stated as a main cause of inaccurate (biased) forecasts in 25% of projects, whereas this was the case for zero road projects.[44] This does not mean that estimates of costs and benefits of planned roads are never deceptive. This source of bias, however, appears to be less prevalent and less systematic for road than for rail projects.   Finally, road projects are more common than rail projects, so the opportunity for learning is greater as well.[45]

For the largest public infrastructure projects, such as the Sydney Opera House, the Channel tunnel, concert halls or stadiums, the cost blowouts appear to be more heavily weighted towards deception. This suggests new governance procedures are needed to minimize waste in the upcoming infrastructure boom.

------- ------ --------

Add Figure 3 here

------ ------- --------

# Overcoming Delusion and Deception in Large Infrastructure Projects

Delusion and deception are not insurmountable. While every large infrastructure project has its own idiosyncrasies, these projects are all prone to delusion and deception, regardless of the private or public funding institution. This section of the article addresses some prescriptive governance advice for overcoming delusion and deception. First, we focus on possible techniques useful for overcoming deception. We address financial and non-financial incentives[46] for the agents, namely the proposing and the approving institutions, planners and contractors. Then, we discuss the adoption of an outside view, which deliberately avoids the details of the case at hand and simply focuses on understanding the historical statistics and patterns of similar projects. The specific outside view tool we use is reference class forecasting, where decision makers search for an unbiased, representative population of similar and past cases that will become useful to make unbiased predictions of the future. When delusion and deception are difficult to disentangle, reference class forecasting has been successful in overcoming both dysfunctional behaviors in diverse settings, including major transport scheme forecasting[47] and movie forecasting.[48]

## Overcoming Deception through Accountability and Transparency

Financial and non-financial incentives should be given to the agents of taxpayers and the institution proposing the project. We first discuss incentives for proposing and approving institutions and then we turn our attention to the incentives for planners and bidders.

### Incentives for the Agents of the Taxpayers: Institutions Proposing and Approving Large Capital Infrastructure Projects

As shown by Flyvbjerg,[49] artificially low costs, exaggerated benefits and underestimated risks are common strategies employed by the proposing institution to have a large infrastructure project approved. This is facilitated by the asymmetric information existing between those who propose a large infrastructure project and those who approve and fund it. In addition, lack of clear accountability and the misalignment of time horizons may lead the proposing individual(s) to take more risk than the funding institution or the taxpayers would like. To overcome deception (as well as empire building motives) there are two key best practices that have been employed: (1) the proposing and the approving institutions should share financial responsibility and (2) private financers should participate in financing the project with their own capital at risk.[50] We examine these issues in turn.

Institutions proposing and approving large infrastructure projects should *share financial responsibility* for covering cost overruns and benefit shortfalls resulting from misrepresentation and bias in forecasting, which helps align incentives. In a recent consultation document,[51] the U.K. Department for Transport proposes a requirement for all large infrastructure projects that asks for funds from the Department to have a minimum local contribution of 10% (25% for light rail) of the gross cost in order to gain program entry, upon the belief that "if an authority has a financial stake in a scheme this provides a clear incentive to ensure that the right structures and resources are in place to bring it to fruition to time and budget".[52] Recognizing that planners are subject to optimism bias, the Department for Transport requires that all requests for funds include an "Optimism Bias Uplift", which is an empirically based adjustment to a project's costs for different percentiles of cost overruns, on the basis of the project type. The uplifts are computed on the basis of actual cost overruns in a reference class of completed projects comparable to the project seeking funding. For example, if the funding institution was prepared to accept a 50% risk of cost overrun (the '50% percentile') on a road scheme then an uplift of 15% should be applied. But if the funding institution was only prepared to accept a 20% risk of cost overrun (the '80% percentile') then a higher uplift of 32% was recommended. In addition, those uplifts are different for project types - the same 50% percentile uplift for rail would be 40% (which is almost three times the amount for roads) – and for different stages of development – uplifts are the highest for "programme entry" and the lowest when the project is receiving "full approval". Notice that these top down estimate uplifts encompass the different complexities of the projects without going into specific details, which is the point of the outside view.

The "Optimism Bias Uplift" is useful to control cost underestimation before the approval of projects. If no measure is taken to control cost escalation after project approval, promoters would simply "postpone" the appearance of costs during the project construction. Therefore, to discourage cost increases during the implementation of the project, the Department for Transport advises that requests for funds should include an additional risk allowance in the amount of 50% of the Optimism Bias uplift. If during the implementation phase a project requires further expenditures that are within the risk allowance, these do not require additional approval from the funding institution. However, the local authority is expected to contribute at least 50% of the cost increase. The local authority should be expected to fund any expenditure in excess of that risk allowance.[53] In this way a clear incentive has been imposed on the local authority to avoid cost overruns. Such an incentive did not exist before.

Beside the provision of the Optimism Bias uplifts, the UK Department for Transport requires promoters to construct a comprehensive Risk Register to mitigate the risk involved in the implementation of large schemes. This register lists the risks that are likely to affect the delivery and operation of the proposed infrastructure. Construction risks (e.g., timescale and cost perspectives) and operational risks (e.g., maintenance risk and revenue risk) and a share of risks associated with climate change should be included in the register. In addition, it advises that the Risk Register "needs to identify who owns the identified risk. For example, some risks may be transferable through insurance or financial instruments".[54]

In addition, to obtain more realistic forecasts and reduced risk, full public financing or full financing with a sovereign guarantee should be avoided. Whenever possible, the decision to go ahead with a project should be made contingent on the willingness of private financiers to participate *without a sovereign guarantee for at least one third of the total capital needs*. The lower limit of a one-third share of private risk capital is based on practical experience.[55] Contracts should be written in such a manner that risk allocation is balanced, i.e., the risk to private financiers must be real, with no comfy escape clauses that return risk to the taxpayer when things get difficult. This has been beneficial in situations where the project passed the market test as well as when it did not (i.e., whether the project has been subsidized or not). Private lenders, shareholders, and stock market analysts should produce their own forecasts or should critically monitor existing ones. If they were wrong about the forecasts, they and their organizations should be held responsible for the mistakes.

**Incentives for the Agents of the Proposing Institution: Planners and Bidders**

Incentives for planners and bidders encompass two measures that address strategic misrepresentation in the proposal and the bidding phases respectively. We discuss them in turn.

To decrease the likelihood of strategic misrepresentation of costs, timeframe and benefits, incentives aiming to achieve higher transparency should be used in order to align the incentives of the planners (and the proposing institution) to provide more accurate forecasts. To provide incentives to planners to disclose their information regarding the specifics of the projects, *rewards* and *higher criticisms* of the forecasts could be two fruitful alternatives. For instance, financial and non-financial rewards should be promoted for planners who provide realistic estimates. In addition, forecasts should be subject to detailed assessment and criticism, such as expert and independent peer reviews, for projects with major public funding carried out by national or state accounting and auditing offices, like the General Accounting Office in the US or the National Audit Office in the UK, and for projects with private funding by independent private auditors. Other forms of scrutiny, such as public hearings and presentations of the forecasts to the scientific community, should also be encouraged. In the most egregious instances, criminal penalties for seriously misleading forecasts may be warranted.[56]

However, these measures are not sufficient because they do not address opportunistic behaviors that are possible in the bidding phase. In the case of the Sydney Opera House, the Sydney Opera House Act was approved in 1960 with the provision that every 10% increase in the budget would require the Act to be amended by Parliament. This did not hold promoters, bidders and contractors from supporting underestimated costs and time for completion. In fact, during the tender, bidders can act opportunistically by assessing the probability that compensation is possible after the construction stage has been initiated. If compensation is possible, bidders will bid the lowest possible value in order to win the tender. The winning bidder will be typically the bidder who most underestimates the true costs of the project. We call this, the "winner's blessing".[57] After the project bas been initiated, the initial low price will be compensated through overpricing the expected scope increases, which the experienced bidders know are almost certain. When compensation is not possible, there is less chance that the bidding price is artificially low.

In this situation, the incentive is to *place financial risk with bidders*. In doing so, bidders have the incentive to disclose any specific information they have regarding costs and completion times, which is often a source of asymmetric information between bidders and the institution managing the construction of the infrastructure. If appropriate measures to overcome deception are taken in the bidding stage, the construction stage should go rather smoothly. However, *placing financial risk with contractors* for delays and scope increases should be a safeguard to be used in conjunction with other measures, especially with regard to coordination between contractors and between contractors and the client.

So far, we have illustrated several contractual arrangements that can be used to decrease the likelihood of strategic misrepresentation of costs, timeframe and benefits by transferring risks to the entities that would otherwise benefit from such misrepresentations. Rather than being part of separated arrangements, these measures can be *bundled* into one contract and a private sector entity can be charged with providing a flow of infrastructure services over time that goes beyond the provision of the building. For example, with a Design-Build-Finance-Operate-Maintain (DBFOM) contract,[58] the private sector entity is responsible for the design, building, financing, operation and maintenance of an infrastructure under a very long period of time, usually 20-30 years, after which the facility is transferred to the public entity. This type of contract addresses in one place several of the mis-alignment issues discussed in this article. These are relatively new mechanisms and deserve further attention.

Table 1 summarizes the key causes of deception and the proposed prescriptive advice.

------- ------ --------

Add Table 1 here

------ ------- --------

**Overcoming Delusion and Deception through the "Outside View"**

There are several instances where delusion and deception cannot be disentangled. To overcome both of them, the key recommendation is to adopt the *outside view* and, in particular, a forecasting method called "reference class forecasting".[59]

The inside view is the conventional and intuitive approach in planning new projects. The traditional way to think about a project is to focus on the project itself and its details, paying special attention to its unique or unusual characteristics, and trying to predict the events (e.g. strikes or weather) that could influence its success. It essentially ignores the details of the case at hand, and involves no attempt at detailed forecasting of the future history of the project. Instead, it focuses on the statistics of a class of cases chosen to be similar in relevant respects to the present one. For example, similarity could be determined by project type, governance structure, complexity, etc. The case at hand is also compared to other members of the class, in attempt to assess its position in the distribution of outcomes for the class.[60] Using the outside view executives and forecasters are not required to make scenarios, imagine events, or gauge their own and others' levels of ability and control, so they do not risk mis-estimating these factors.

When both the inside and the outside view of forecasting are applied with equal skill, the outside view is much more likely to produce a realistic estimate.[61] In very few instances, since it is based on historical precedent, the outside view may fail to predict extreme outcomes such as those that lie outside all historical precedents. But for most projects, the outside view will produce more accurate results.

The outside view can be implemented through a technique that has been defined as reference class forecasting. It requires the decision maker to obtain a reference class of past, comparable cases when making predictions about costs and benefits of a new project. By introducing distributional information of successful as well as unsuccessful past projects, the decision maker is forced to consider the entire distribution of possible outcomes. This prevents the decision maker from focusing on easily recalled similar projects, which are typically successful ones.[62] The implementation of reference class forecasting is organized into five steps.[63]

(1) **Select a reference class**. Identifying the right reference class involves both art and science. The decision maker usually has to weigh similarities and differences on many variables and determine which are the most meaningful in judging how the project at hand will play out. Sometimes that is easy. A planner who has to forecast the construction costs of a rail project planned in the manner that such a project is usually planned around the world would easily find a reference class. In other cases, especially when the project requires the incorporation of a new technology, it is more difficult. The key is to choose a class that is broad enough to be statistically meaningful but narrow enough to be truly comparable to the project at hand.

(2) **Assess the distribution of outcomes**. Once the reference class is chosen, the decision maker has to document the outcomes – in terms of whichever variable is considered pertinent, e.g. cost overrun, total costs or unit costs – of the prior projects and arrange them along a distribution of outcomes, showing the extremes, the median, and any clusters. Sometimes it will not be possible to precisely document the outcomes of every member of the class. However, a rough distribution can still be obtained by calculating the average outcome as well as a measure of variability. Obtaining good projects with valid data is the hardest and most time-consuming part of reference class forecasting. Results will only be as good as the projects and data that are used as input to the exercise. As part of data validation it must be decided what type of issues data take into account, e.g., uncertainty, design changes, and the like.

**(3) Make an intuitive prediction of your project's position in the distribution**. Based on his or her own understanding of the project at hand and how it compares with the projects in the reference class, the decision maker needs to predict where it would fall along the distribution. Because the intuitive estimate will likely be biased, the final two steps are intended to adjust the estimate in order to arrive at a more accurate forecast.

**(4) Assess the reliability of your prediction**. This step is intended to gauge the reliability of the forecast made in Step 3. The goal is to estimate the correlation between the forecast and the actual outcome, expressed as a coefficient between 0 and 1, where 0 indicates no correlation and 1 indicates complete correlation. In the best case, information will be available on how well the decision maker's past predictions matched the actual outcomes. Then, the correlation based on historical precedent can be estimated. In the absence of such information, assessments of predictability become more subjective. An estimate of predictability may be based on how the situation at hand compares with other forecasting situations. Through a diligent statistical analysis, the decision maker could construct a rough scale of predictability based on computed correlations between predictions and outcomes for other endeavors like road or bridge construction. He or she can then estimate where his or her ability to predict rail project construction costs lies on this scale. When the calculations are complex, it may help to bring in a skilled statistician.

**(5) Correct the intuitive estimates.** Due to the bias, the intuitive estimate made in Step 3 will likely be optimistic – deviating too far from the average outcome of the reference class. In this final step, the estimate is adjusted toward the average based on the analysis of predictability in Step 4. The less reliable the prediction, the more estimate needs to be regressed toward the mean. Suppose that the intuitive prediction of the construction costs is $4 billion and that, on average, rail

projects in the reference class cost $7 billion. Suppose further that the correlation coefficient has been estimated to be 0.6. The regressed estimate of construction costs would be:

$$\$7B + [0.6 (\$4B - \$7B)] = \$5.2B$$

Thus, the adjustment for optimism will be substantial, particularly in highly uncertain situations where predictions are unreliable.

Besides overcoming optimism bias and mitigating the problems that derive from strategic misrepresentation, reference class forecasting provides two other major benefits. First, since the reference class comprises previous projects, it helps to look across types of projects, geographies and types of financing methods to provide much more information to the decision maker. This represents a test to see which projects have worked out in the past and the kind of tweaks to the model that might increase the chance that the project at hand would be a success. Second, it helps to provide a reality check on whether the project is likely to perform up to expectations.

The first instance of reference class forecasting in practice was carried out in 2004 estimating the costs of the proposed Line 2 of the Edinburgh Tram. The business case estimations prepared by the promoters included a base cost of £255 million (US$400 million) and an allowance for covering contingency and optimism bias that amounted to 25% above the base cost. However, a reference class of 46 comparable rail projects indicated that the promoters' estimates were optimistic and that optimism bias uplifts should be applied. The reference class was established by Bent Flyvbjerg and Cowi in close collaboration with experts at the UK Department for Transport.[64] Rail projects from Flyvbjerg's megaproject database were screened for inclusion, focusing on projects that had been planned and built under comparable regulatory and contractual regimes. Statistical tests were applied to decide whether projects were indeed comparable, before

inclusion in the reference class. After this, an independent review applied optimism bias uplifts to the promoters' capital cost estimates, based on cost overruns in the reference class as required by the UK Department for Transport, and thereby taking an outside view on the cost forecasts. The review concluded that total capital costs were more likely to be £357 million, with a 50% risk of going over budget. If the client wanted to reduce the risk of going over budget to 20%, higher uplifts had to be used and the capital cost budget would have been £400 million.[65]

The resulting reaction to the reference class forecast was that the proposed tram was re-evaluated. Scottish Finance Secretary John Swinney publicly stated, "I want to be absolutely sure about the calculation of the costs involved in these projects, and the assessment of risk involved, before they progress further" (*The Scotsman*, June 6, 2007).

The benefits of using reference class forecasting are present both when delusion is substantial (and deception is relatively less important) and when deception is substantial (and delusion is relatively less relevant). The key is to choose a similar class of situations to the focal problem. In the first instance, cost and benefit estimates are affected by optimism. Reference class forecasting bypasses bottom up estimates for the project at hand and uses the actual outcomes (not biased estimates) of similar past projects. In the second instance, these estimates are not affected by optimism but have been purposely misrepresented to get the project approved. By using realized outcomes of similar past projects rather than manipulated estimates of the current project, reference class forecasting provides more reliable, top down estimates of the true costs and benefits of the project. Reference class forecasting helps both to avoid common cognitive biases and strategic manipulation in order to produce more accurate forecasts. A new and related forecasting technique called Similarity Based Forecasting may provide even more accurate forecasts but has yet to be proved on infrastructure projects.[66]

# The Way to Go

Large infrastructure investments are a vital component of any public or private institution. Unfortunately, cost overruns, delays and exaggerated benefits are the norm rather than the exception for roads, bridges, stadiums, concert halls, new plants, etc.

Although large infrastructure projects occur frequently across the globe, any individual project is often a once in a career decision for a public or private executive. Thus it is difficult for executives to learn from their own prior mistakes. It is rare for executives to deliberately learn from similar projects other have attempted. Typically executives to adopt an *inside view* of any particular problem – where they focus on the specifics of the case at hand. Without the opportunity to learn from rapid and unambiguous feedback regarding their estimates of costs and benefits, executives can hardly learn from these unique decisions and avoid making similar mistakes in future projects. In such situations, inside view thinking leads to numerous cognitive biases that result in optimistic delusions.

These, often individual, optimistic delusions are confounded, sometimes even dwarfed, by the magnitude of strategic deceptions among the different actors in the system. On several occasions, however, decision makers have attempted to justify their deceptive behavior by arguing that the decision was in the *public interest*. On one hand, it can be argued that public-sector executives may decide to deliberately underestimate costs in order to provide public officials with an incentive to cut costs and thereby to save the public's money. According to this type of explanation, higher cost estimates would be an incentive for wasteful contractors to spend more of the taxpayer's money. Empirical studies have identified executives and planners who say they deliberately underestimate costs in this manner to save public money.[67] Merewitz endorsed and summarized this viewpoint as "keeping costs low is more important than estimating costs correctly".[68]

On the other hand, a second explanation in terms of public interest covers the not uncommon situation where project promoters believe their venture will benefit society and posterity. They feel that they should do anything possible to make the project happen, including cooking forecasts of costs and benefits. Both types of public-interest explanations see the end (project approval) as justifying the means (estimates of costs and benefits that show the project should be approved).[69]

However, these arguments overlook an important fact. Underestimating the costs and overestimating the benefits of a given project results in artificially high benefit-cost ratio, which in turn leads to two problems. First, the project may be started despite the fact that it is not economically viable. Second, a project may be started instead of another project that would have yielded higher returns had the actual costs and benefits of both projects been known. Thus, for reasons of economic efficiency alone, the argument that cost underestimation saves money must be rejected.[70] As a case in point, an ex post benefit-cost analysis of the Channel tunnel between France and the UK showed that the actual net present value of the project to the British economy was minus US$17.8 billion and the actual internal rate of return minus 14.45 percent. The study concluded that "The British Economy would have been better off had the Tunnel never been constructed".[71]

Because delusion is often accompanied by strategic deception, this study's prescriptive advice has been broken into two parts. First, we focused on best practices to diminish strategic deceptions (e.g. P-A issues) in the specific context of infrastructure projects. Next, we examined how executives can adopt an "outside view" of problems by using reference class forecasting. This statistical procedure uses both a forecaster's intuition and historical data to mitigate the two types of errors and arrive at a more accurate estimate. The American Planning Association has recommended this procedure for large infrastructure projects. Its widespread use would surely produce more accurate estimates of large infrastructure projects and projects like Toll Collect and the Channel tunnel would be profitably and happily foregone by the vast majority of the public. Ultimately, accurate reference class forecasting, proper incentives and budgets are the way to go.

# Notes

**Figure 1: Pioneer process plants cost forecasts accuracy (Merrow, Phillips and Meyers, 1981)**

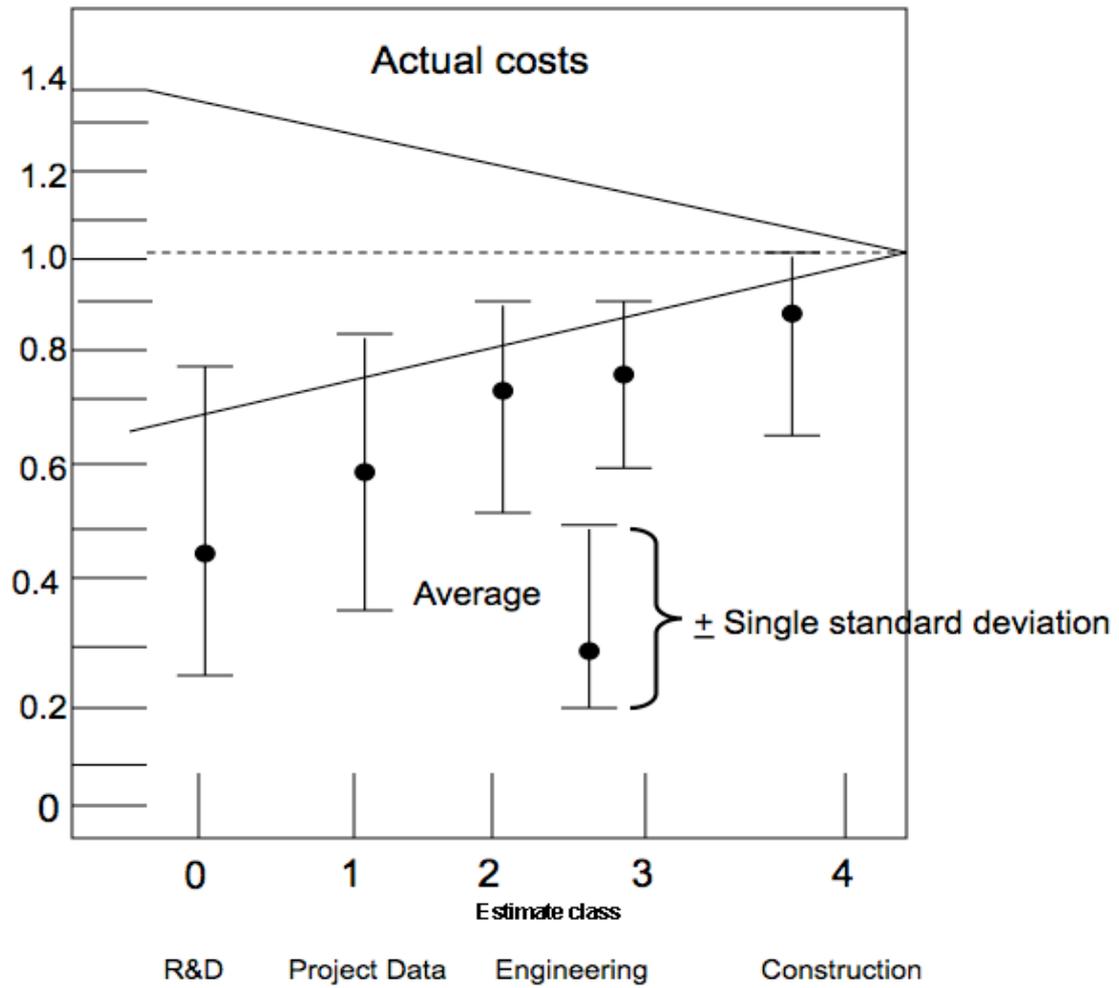

**Figure 2: Multi-tier principal-agent relationships**

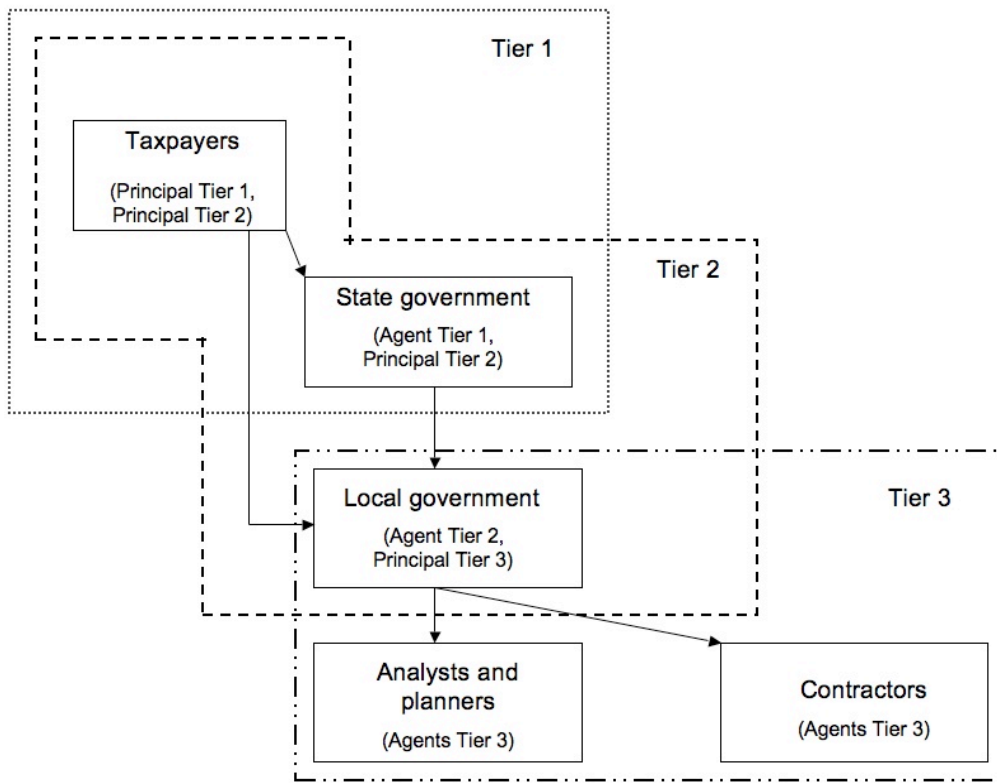

**Figure 3: Incentives and learning in large capital investments**

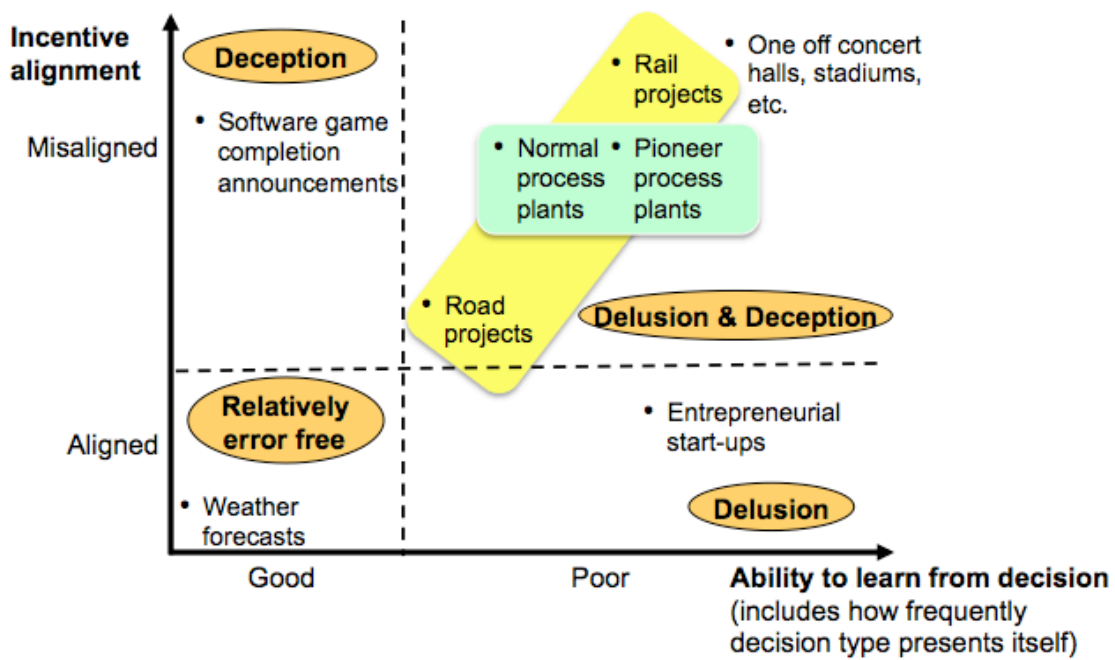

**Table 1: Avoiding strategic deception**

| Actors | Causes of deception | Prescriptive advice |
|---|---|---|
| 1. Proposing and approving institutions (i.e. agents of the taxpayers) | Proposing institution strategically misrepresents costs, timeframe, risks and benefits to obtain funding | Proposing and approving institutions should share financial responsibility (e.g. minimum local contribution, local contribution for cost increases, identification of who "owns" the risks)<br><br>Private financers should participate, without a sovereign guarantee, for a least one third of the total capital needs |
| 2. Planners, bidders and contractors (i.e. agents of the proposing institution) | Planners strategically misrepresent costs, timeframe and benefits (to please the proposing institution) | Financial and non-financial rewards for planners who proposed realistic estimates<br><br>Strict forecasts audit<br><br>Criminal penalties for purposely misleading forecasts |
| | Bidders propose artificially low bids because of planned compensation through expected scope increases | Place financial responsibility with bidders |
| | Contractors overprice scope increases | Place financial responsibility with contractors for delays and scope increases |